\documentclass{aa}
\usepackage{graphicx}
\begin{document}
\title{Cluster formation versus star formation rates around six
regions in the Large Magellanic Cloud}
\author{Annapurni Subramaniam}
\institute{Indian Institute of Astrophysics, II Block, Koramangala, Bangalore
560034, India}
\offprints{Annapurni Subramaniam, e-mail: purni@iiap.ernet.in}

\date{Received / accepted}

\authorrunning{Subramaniam}
\titlerunning{Cluster and star formation rates in LMC}

\abstract{
The stellar population and star clusters around six regions in the
Large Magellanic Cloud (LMC) is studied to understand the correlation between
star formation and cluster formation rates.
We used the stellar data base of the OGLE II LMC survey and the star
cluster catalogues. The observed distribution of stellar density on the
colour-magnitude diagrams (CMDs) were compared with the synthetic ones generated
from the stellar evolutionary models. By minimising the reduced $\chi^2$ values,
the star formation history of the regions were obtained in terms of star formation rates (SFR).
All the regions were found to show large SFRs between the ages 500 -- 2 Gyr with lower values
for younger and older ages. 
The correlated peak in the
cluster and SFRs is found
for ages $\sim$ 1 Gyr, and for ages less than 100 Myr.
Five out of six regions show significant cluster formation in the
100 -- 300 Myr, while the SFRs were found to be very low. This indicates
anti-correlation between star and cluster formation rates for the 100 -- 300 Myr age range.
A possible reason may be, that the stars are predominantly formed in clusters, 
whether bound or unbound, as a result of star formation during the above age range. 
The enhanced cluster formation rate at
 100 -- 300 Myr age range could be correlated with the encounter of 
LMC with the the Small Magellanic Cloud, while the enhanced star and cluster formation 
at $\sim$ 1 Gyr does not correspond
to any interaction. This could indicate that the star formation induced by interactions is
biased towards group or cluster formation of stars.
}

\maketitle

\section{Introduction}

In the recent years, the Large Magellanic Cloud (LMC) has been very thoroughly 
studied using various surveys,
for example, OGLE II (Udalski et al. \cite{u2000}), Magellanic Clouds Photometric Survey 
(Zaritsky et al. \cite{z97}). These surveys were used partly or fully to study the 
star clusters and the stellar population in LMC. However,
the spatial correlation between the star and cluster formation at small scales 
are not studied.  Girardi et al. (\cite{g95}) used the star cluster catalogue
of Bica et al. (\cite{bea96}) to derive the star cluster properties in LMC.
Pietrzynski \& Udalski (\cite{pu00}) used the OGLE II data and studied the age
distribution of LMC star clusters.
There were also studies on star clusters and stellar population around them,
(for example, Olsen et al. \cite{o98} and Olsen \cite{o99}), but these also do not 
compare the spatial correlation between cluster and star formation episodes. 
The recent study by Holtzman et al. (\cite{h99}) suggested that the star formation history
of the field stars is different from that of the clusters. This difference is seen
in the age range 4.0 -- 12 Gyr, where there seems to be a paucity in cluster formation.
It is concluded that in general, the star clusters in LMC are not good tracers
of the stellar population (van den Bergh \cite{v99}). 
This conclusion has been made from the analysis
of the cluster and stellar population in the whole of LMC. 
In this study, an attempt is made
to study the correlation between the star formation and cluster
formation episodes around a few regions in LMC. Also, the emphasis is on the
younger age range, for ages $\le$ 1 Gyr and at smaller scales.

The recent star forming regions, like the 30 Dor and super giant shells found in
LMC indicate
that the star formation which began at one point propagates to larger distances
in the LMC.  The theories put forward to explain these structures include
stochastic self-propagating star formation, SSPSF (Feitzinger et al. \cite{f81}) and
recently by de Boer et al. (\cite{b98}), suggesting bow-shock induced star formation.
The correlation between the events of cluster formation and star formation is expected in
the resulting stellar population. We explore this correlation in this study, by looking
at star clusters surrounding a few regions in LMC.

\section{Data }
\subsection{Regions}
\begin{figure}
\resizebox{\hsize}{!}{\includegraphics{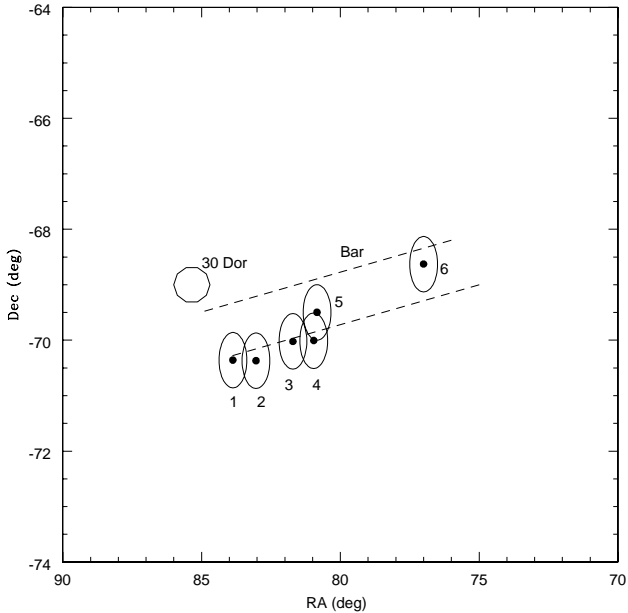}}
\caption{The location of the regions studied are shown on the LMC. The Bar and
30 Dor are also shown.}
\label{figure1}
\end{figure}

The regions studied here were chosen for a different project, 
that is, to study the stellar population around novae in LMC (see
Subramaniam \& Anupama \cite{sa02}). From the above study,
six regions were selected for the present analysis.
The stellar data within a radius of a few arcmin is used to study
the star formation
history (SFH) of the region under consideration, whereas the star clusters 
are identified within 30 arcmin radius to study the
cluster formation events. The above value of 30 arcmin is chosen such that
the area covered is similar to the size of supergiant shells in LMC. 
The regions selected here satisfy the following two conditions - a) there are good
number of star clusters within 30 arcmin radius and b) ages are known 
for most of the identified clusters.
The location of these regions are given in Table 1. These
locations are also plotted in figure ~\ref{figure1}. The dots show the centers
of the regions studied and the big circles around them show the extend of the
region scanned for star clusters.


Field stars within a radius of a few arcmin around these regions were
identified from the OGLE II survey (Udalski et al. \cite{u2000}). 
We used 
the photometric data in the V and B pass bands and V vs (B$-$V) 
colour-magnitude diagrams (CMDs) of the identified field stars were used for
further analyses.  
\begin{table*}
\caption{Location of the six regions in the Large Magellanic Cloud.}
\begin{tabular}{lllllll}
\hline
Region &\multicolumn{3}{c}{RA}&\multicolumn{3}{c}{Dec}\\
     & h & m & s & $^\circ$ & $^\prime$ & $^{\prime\prime}$ \\
\hline
Region 1  & 05&35&29.33& $-70$&21&29.39\\
Region 2  & 05&32&09.27& $-70$&22&11.70\\
Region 3  & 05&26&50.33& $-70$&01&23.08\\
Region 4  & 05&23&50.12& $-70$&00&23.50\\
Region 5  & 05&23&21.82& $-69$&29&48.48\\
Region 6  & 05&08&01.10& $-68$&37&37.67\\
\hline
\end{tabular}
\end{table*}

\subsection{Star clusters}
The star clusters in the vicinity of six regions were identified and their properties
obtained based on the
following catalogues: Pietrzynski et al. (\cite{pea99}) (P99), Bica et al.\
(\cite{bea99}) (B99), Bica et al.\ (\cite{bea96}) (B96), Pietrzynski and Udalski 
(\cite{pu00}) (PU2000). 
B96 presented integrated UBV photometry of 624 star clusters and associations 
in the LMC. They estimated the ages of the clusters based on their integrated 
colours and hence classified the clusters into SWB types (Searle, Wilkinson \&
Bagnoulo \cite{swb80}), which is basically an age sequence. This classification can
be used to obtain the approximate age of the clusters.
B99 is a revised version of the above catalogue and contains about 1808 star
clusters for which the positions and extents are tabulated. 
P99 presented photometric data of 745
star clusters and their nearby field, of which 126 are new findings. 
PU2000 estimated the ages for 600 star clusters
presented in the P99 catalogue. The catalogues in B99 and P99 were used to
identify the clusters, while B96 and PU2000 were used to estimate the ages of 
the identified star clusters.

Clusters have been identified within 30~arcmin radius around 6 regions.
109 clusters have been identified near 6 regions. Of these,
age estimates for 89 clusters could be obtained from PU2000 and B96. 
The B96 gives the age of the cluster in terms of groups. Since the interest is in
age groups of the cluster population rather than the ages of the
individual clusters, the above data serves the purpose. Therefore, even those
clusters whose exact age is known are also grouped. No systematic shift in the
cluster age is found between the two catalogues.
The number of clusters 
detected near each region, the number for which the age is known and the number 
of clusters in various age groups are tabulated in Table 2. 

Field stars within a radius of a few arcmin 
are analysed to study the star formation history, while clusters within 30 
arcmin ($\sim$ 400 pc) radius are considered to identify the cluster formation
episodes.
The choice of larger radius for the clusters is justified as they are being used to study
the star formation events which took place on relatively larger
scales. It is found that the supergiant shell LMC 4 is about 1 Kpc in diameter.  
 The size of the supergiant shell can be considered to be typically the area covered 
by a propagating star formation. Therefore, we have chosen
very similar length scale for identifying the star clusters.
\begin{table*}
\caption{Statistics of star clusters identified near the regions.}
\vspace{0.2cm}
\begin{tabular}{cccrrrrr}
\hline
Region &  No. of clusters & No. of clusters & \multicolumn{5}{c}{Age groups}\\
     &  within 30 arcmin&  with age known & $\le$ 7.5&7.5\,--\,8.0&8.0\,--\,8.5
&8.5\,--\,9.0&$\ge$ 9.0\\
\hline
Region 1&     15 &12 & 1 & - & 7 & 3 & 1 \\
Region 2&      9 & 8 & - & 4 & 1 & 2 & 1 \\
Region 3&     17 &13 & - & 3 & 3 & 3 & 4 \\
Region 4&     24 &19 & - & 2 &11 & 4 & 2 \\
Region 5&     17 &15 & 1 & 1 & 6 & 5 & 2 \\
Region 6&     27 &22 & - & 4 &11 & 7 & - \\
\hline
\end{tabular}
\end{table*}

\begin{figure*}
\centering
\includegraphics[width=17cm]{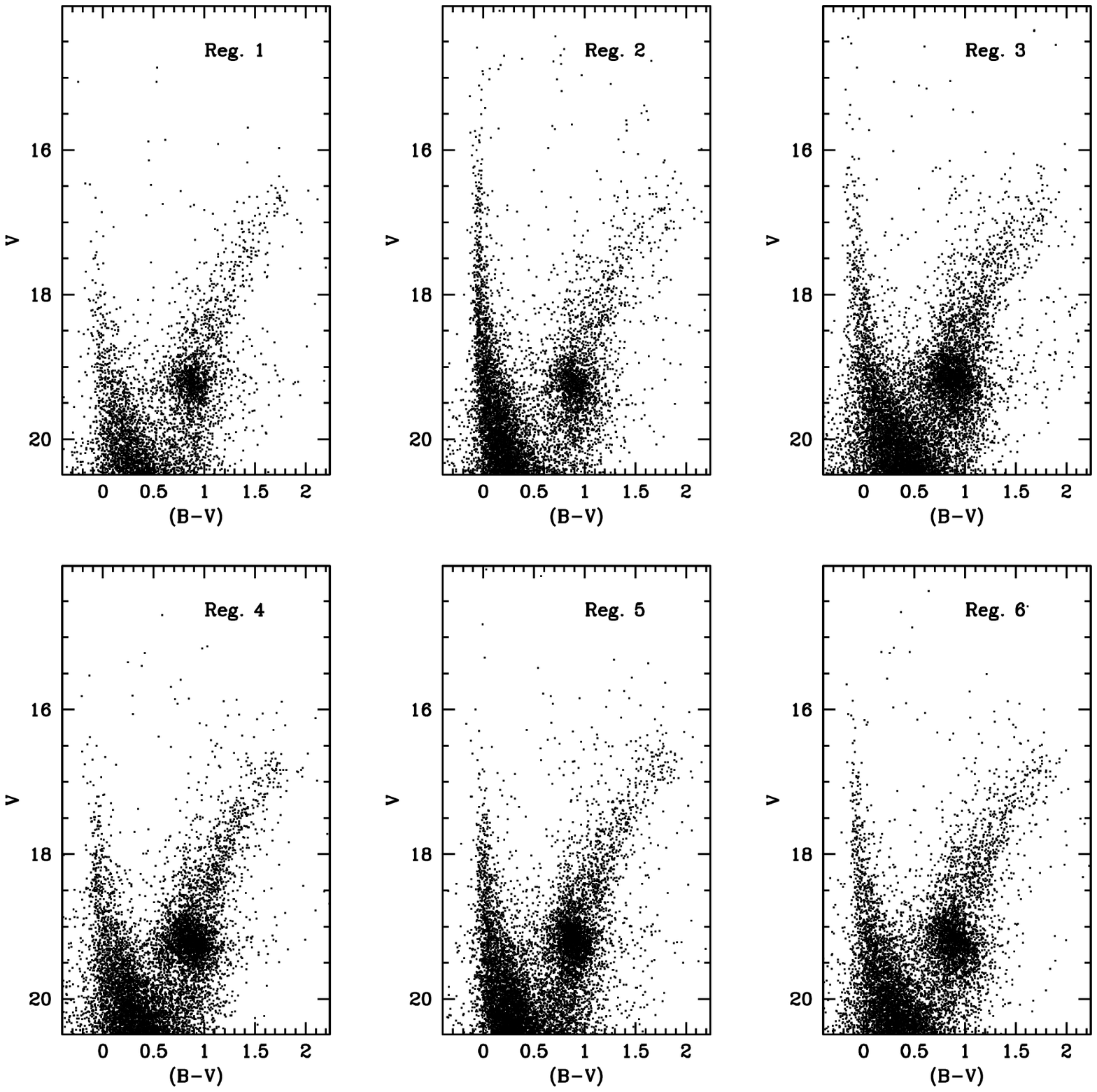}
\caption{The V vs (B$-$V) CMDs of all the six regions studied here are shown.
These CMDs are used to estimate the stellar density after correcting for data
incompleteness.}
\label{figure2}
\end{figure*}

\section{Estimation of star formation history}
The star formation history of the regions studied are derived using the CMDs.
The CMDs represent  stellar population of different ages present in the region.
The brightest main-sequence (MS) stars represent the stars born out of the recent
star formation event and the red giants represent stars born from an earlier star
formation events. We use the technique of synthetic CMDs to reproduce the
observed CMD and thereby estimate the star formation history.

\subsection{Observed CMDs}
The observed V vs (B$-$V) CMDs are obtained for stars located within a radius of a few
arcmin from the center. The CMDs are shown in figure \ref{figure2}.
The CMDs are affected by the reddening towards the 
observed region, photometric incompleteness in the data which is a function of
stellar crowding and stellar magnitudes, and the distance.
The incompleteness in the data is tabulated in Udalski et al. (\cite{u2000}) and these
values are used for correcting the incompleteness. 
The CMD is made into a two dimensional array
by binning in V and (B$-$V). The width of the bin in V is 0.2 mag and (B$-$V) is 0.1 mag.
The number of stars in each bin is counted and then corrected for the incompleteness.
For each box in the CMD, there would be separate values for the data incompleteness
corresponding to the V and B frames, and the smaller value was used for the correction. 
After correcting for the incompleteness, 
two-dimensional stellar count was obtained. The number density of stars in each box is 
estimated by normalising the above values with respect to the total
number of stars in the CMD. This density distribution of stars in the CMDs
is used to compare with the synthetic CMDs.
 The selection of the values of reddening and
the distance modulus are discussed section (3.3).

\subsection{Synthetic CMDs}
The synthetic CMDs are made with the help of stellar evolutionary models after
assuming a model for the age-metallicity relation. We used the evolutionary models
of the Padova group, Fagotto et al. (\cite{f94a, f94b}). The following relation
between the age and metallicity was found to be satisfactory, for ages less than
4 Gyr, we used the Z=0.008 models, for ages between 4 and 9 Gyr, 
we used Z=0.004 models and for ages more than 9 Gyr and upto 12 Gyr, we used
Z=0.0004 models. 
A small fraction of binaries ($\sim$ 10\%) and the photometric error
in the observation are also included in the synthetic CMDs. The algorithm used here
is an extension of that used in Subramaniam \& Sagar (\cite{ss95, ss99}). In this algorithm,
we introduced the age range such that we obtain heterogeneous population in age.


The procedure is described below. The synthetic CMDs for a set of ages with small age ranges were
created. 
The stellar distribution in the synthetic CMDs are converted to density distribution, 
using a procedure
similar to the observed CMD. These synthetic CMDs are used as templates of the stellar density 
distribution for various age ranges. The synthetic CMDs were created 
with sufficiently high number of stars in order to minimise the statistical 
fluctuations. The above mentioned templates are created for 10 -- 50 Myr, 51 -- 100 Myr and then
for a gap of 100 Myr till 1000 Myr, thereby creating 11 templates. For the 50 Myr
age range, the stars were created in a step size of 5 Myr, whereas for 100 Myr
age range, a step size of 10 Myr was used.  For ages beyond
1 Gyr, the templates were created for an age range of 200 Myr, till 2 Gyr, such that 
5 templates were made available. Between 2 and 5 Gyr, 6 templates were created with an age 
range of 500 Myr. Though the present data is not very suitable for estimating the star
formation history beyond 5 Gyr, for the sake of completion, the templates for ages older than
5 Gyr is also included. For older ages, the templates were made for 5 -- 6 Gyr, 6 -- 8 Gyr,
8 -- 10 Gyr and 10 -- 12 Gyr.
These templates are used to create the final CMD in terms of stellar density, which in turn are
compared with the density of the observed CMD. The density of the templates
are scaled and combined to obtain the best fitting synthetic CMD, based on the $\chi^2$ minimisation
technique. Scaling the stellar density in various templates is equivalent to adjusting the
star formation rates (SFRs) in the respective ages. The SFRs are estimated in units of
1 X 10$^{-5} M_\odot yr^{-1}$, such that this is the minimum detectable value.
The range in the scaling factor for which the minimum value
of the $\chi^2$ obtained is estimated for each template. The average and the 
deviation about the mean of the star formation rate for each age range are thus estimated.

The technique used here is very similar to that used in Dolphin (\cite{d97}), 
Olsen (\cite{o99}) and Dolphin (\cite{d00}). 
For stars younger than 1 Gyr the resolution in age is more, whereas it is less 
for population older than 1 Gyr. As we attempt to
compare the cluster formation episodes, which have higher resolution at younger ages,
the above values of resolution are adequate.
The limiting magnitude in the OGLE II data is around $V = 21.0$~mag. This implies
the stars in the MS are younger than about 1.6 Gyr, while the RGB stars are
a mixture of both young and old population. Therefore the present data is not suitable
to understand the star formation history older than 4 Gyr. Also the cluster ages
are known to have a gap between 4 and 10 Gyr. Although we have considered templates upto
12 Gyr, the comparison between the star and the cluster formation episodes stop at
4 Gyr. 

\subsection {Reddening and distance modulus}
The reddening towards each region is estimated by comparing the location of the
observed MS stars with the MS of the templates younger than 1 Gyr. The comparison
was made with the stellar CMD.
The regions 1, 3 and 4 were found to have a reddening of E(B$-$V) = 0.06 mag,
regions 2 and 5 were found to have E(B$-$V) = 0.12 mag and the region 6 was found to have
E(B$-$V) = 0.08 mag. The reddening estimates have an error of 0.02 mag. These
values of reddening agree very well with the reddening estimates as  found by
Subramaniam (\cite{as03}) for various location in the bar region, which is similar to the
estimates by Udalski et al. (\cite{u99}), but with better spatial resolution. 
These values are also consistent with the values previously estimated 
for different location in the LMC and agrees (within errors) with 
those estimated by Dolphin (\cite{d00}). Following Pietrzynski \& Udalski (\cite{pu00}) a 
distance modulus of 18.24 mag for the LMC is assumed initially.
While creating the
synthetic CMDs, the value of the distance modulus was changed and the value giving rise
to the minimum value of $\chi^2$ is adopted. A value of 18.30 $\pm$ 0.06 was found to
produce the best fit for all the six regions. 
This would indicate a distance of 45.7 $\pm$ 1.3 Kpc to the LMC.
For this value of distance modulus, 1 arcmin corresponds to 13.4 pc on LMC. 
Thus a region with 30 arcmin radius would correspond to 400 pc.
Thus the region scanned for clusters is 800 pc in diameter, this is similar to
the sizes of supergiant shells in the LMC. 
 
\section{Results}
\subsection{Region 1}
\begin{figure*}
\centering
\includegraphics[width=17cm]{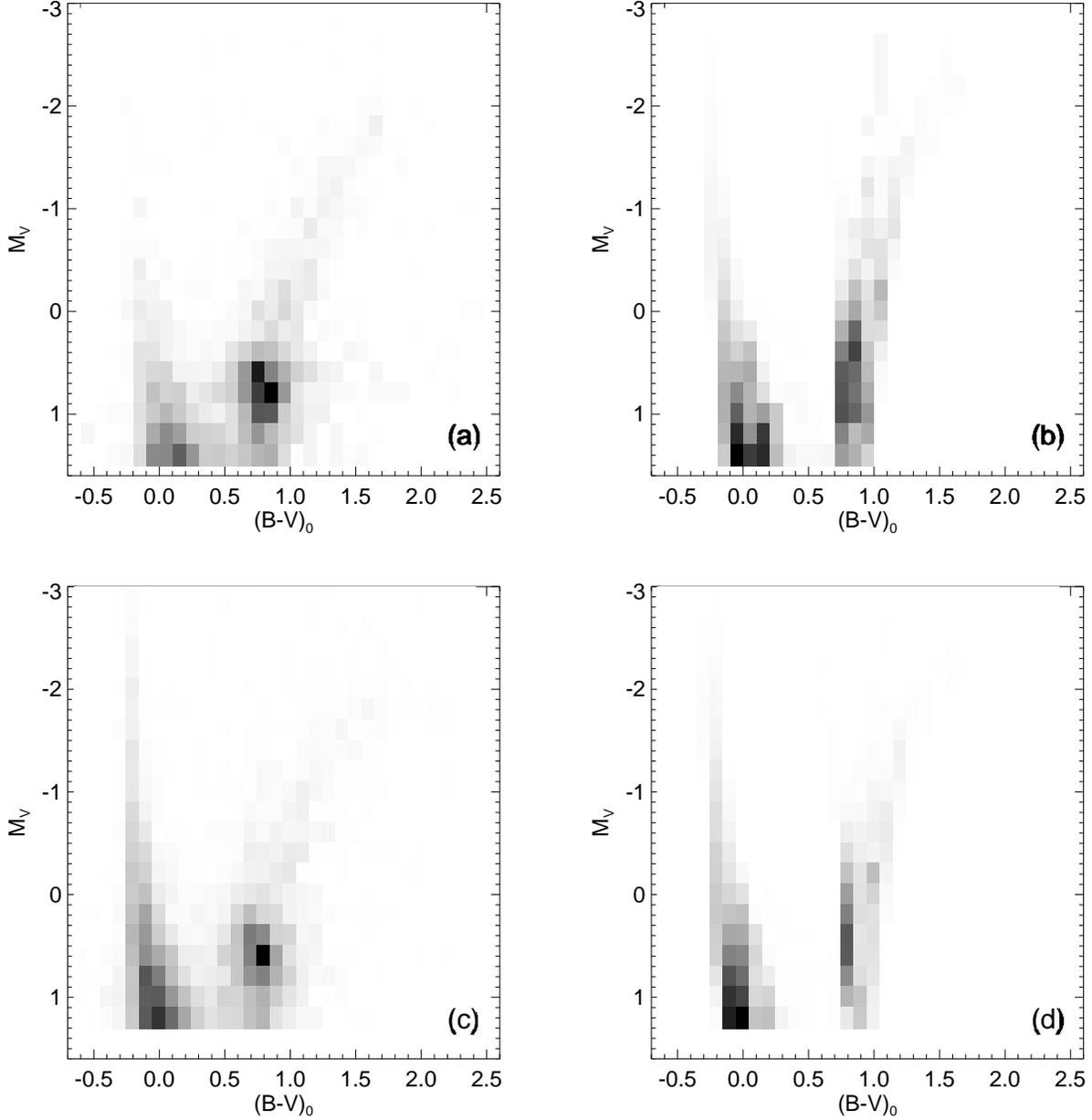}
\caption{The stellar densities on the CMDs for the regions 1 and 2 are shown here. 
The figure (a) and (c) correspond to the observed CMDs, (b) and (d) correspond to the synthetic CMDs
for the regions 1 and 2 respectively}
\label{figure2a}
\end{figure*}
\begin{figure*}
\centering
\includegraphics[width=17cm]{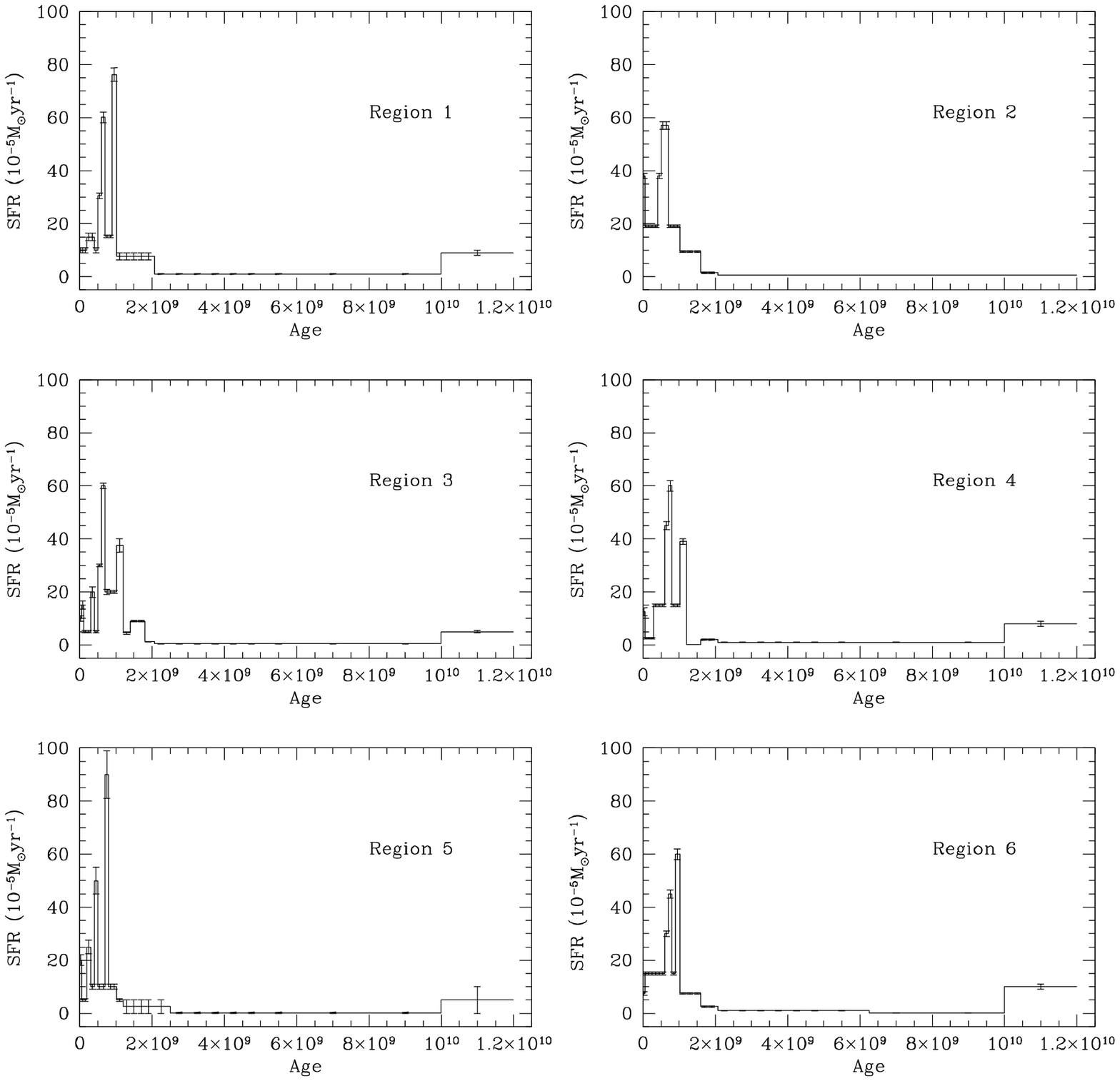}
\caption{The estimates SFRs in the six regions between 10 Myr to 12 Gyr are plotted as
histograms.}
\label{figure2d}
\end{figure*}
\begin{figure*}
\centering
\includegraphics[width=17cm]{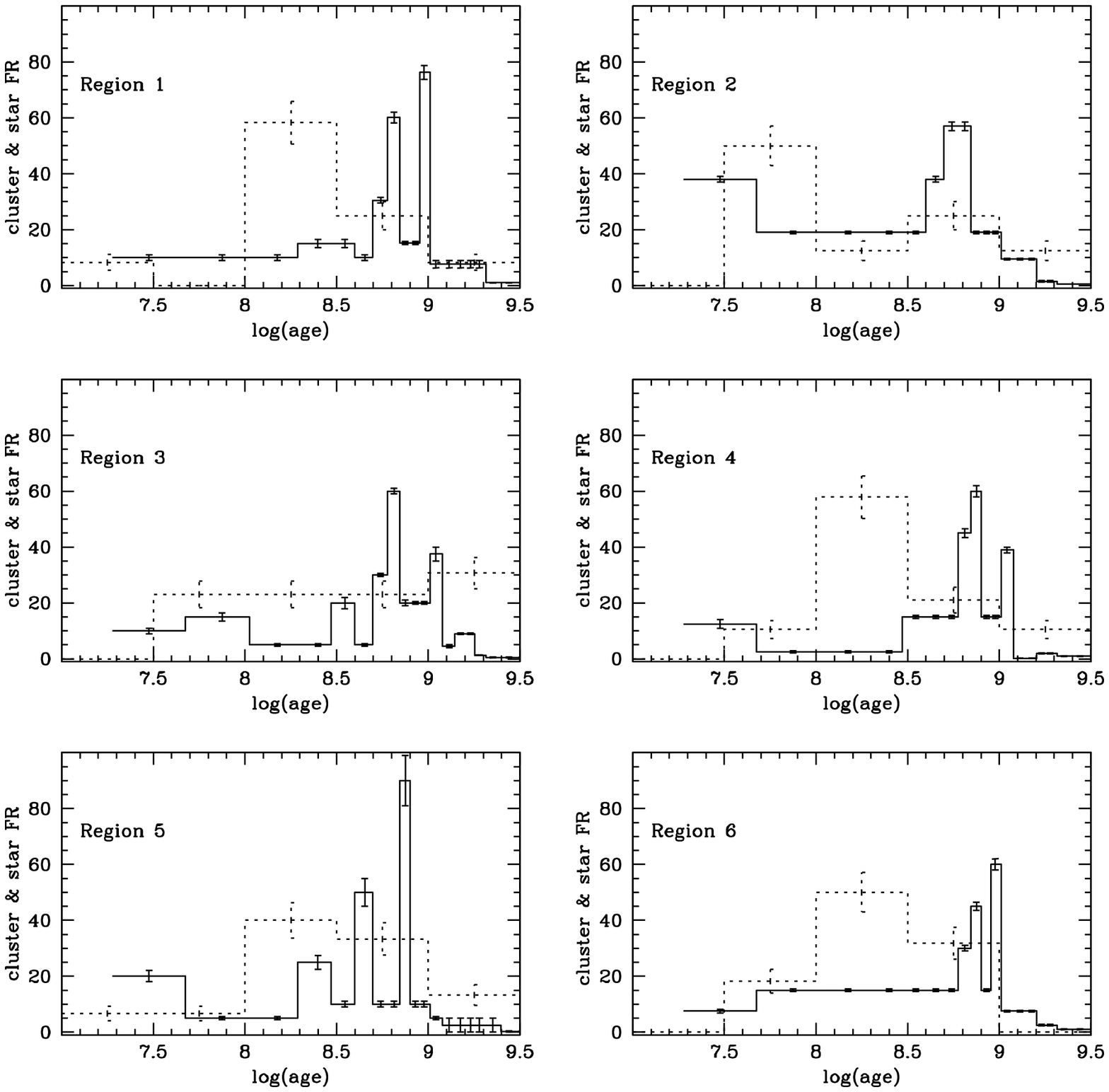}
\caption{The SFRs and the fraction of clusters in various age groups are plotted here
for six regions. This plot helps to find out any correlation between the increaments or
decreaments in the rates of star and cluster formation.
}
\label{figure2e}
\end{figure*}

15 star 
clusters were identified within a radius of 30 arcmin ($\sim$ 400 pc), 
of which the ages are known for 12 clusters. We find that,
58\% of the clusters in this region have ages between 100\,--\,300 Myr, 25\% 
between 300 Myr and 1 Gyr. Only 8\% of clusters are either younger than 30 Myr or
older than 1 Gyr. Hence the bulk of cluster formation has occurred in the 
100-300 Myr range with a tapering towards 1 Gyr. 
Figure~\ref{figure2e} shows the histogram of 
the normalised fraction of clusters with respect to age. 
The CMD of this field consists of 4132 stars
within a radius of 5 arcmin as shown in figure~\ref{figure2}.
The observed stellar density is shown in figure~\ref{figure2a}, along with the synthetic
one. 
It can be seen that the synthetic CMD looks very much like the observed one. The 
part which is not reproduced is the very high density of the red clump stars
in the CMD. 
The other minor effect is the marginally more number of stars seen at the faint
end of the MS. The overall agreement is quite good and the reduced $\chi^2$ value of the fit is
0.07. The estimated star formation rates is shown in figure~\ref{figure2d}.
The region is found to have had a higher rate of star formation between 500 -- 1000 Myr,
then a reduced star formation till 2 Gyr. Between 2 Gyr to 10 Gyr, there was star formation in
a very low rate. This region shows the presence of a burst of star formation
between 10 -- 12 Gyr. Also, we see that this region did not form much stars in the last 500 Myr.

Figure~\ref{figure2e} shows the comparison of SFRs and fraction of clusters formed
in various age groups.
It is seen  that the star formation event between 1.0 and 2.0 Gyr 
resulted in the formation of one
star cluster. The star formation which continued to younger ages,
 resulted in the formation of 3
star clusters, which fall in the age range 300 \,--\, 1 Gyr. 
The enhancement in the
cluster formation between 100 Myr \,--\, 300 yr is not reflected in the formation
of field stars. For ages older than 300 Myr, 
the cluster and star formation events are correlated such that the number of clusters found
between the age group 1 -- 2 Gyr is less than that found in the group 300 Myr  -- 1 Gyr.
On the other hand an anti-correlation is seen for ages younger than 300 Myr.
Also there is very less star formation in the last 100 Myr, whereas there are at least two
star clusters younger than 100 Myr. 
The point to be noted here is that the cluster formation seems to have continued to very
recent times, whereas low SFRs are found in the last 200 Myr. In particular, the
star formation and the cluster formation seems to differ in the last 200 -- 300 Myr history.

\subsection{Region 2}
In this region, 9 star clusters have been identified 
within 30 arcmin radius and the ages of 
8 clusters are known. 50\% of the clusters have ages in the range 30\,--\,100 Myr, 
12.5\% have ages in the range of 100\,--\,300 Myr, 25\% in the range 
300\,--\,1 Gyr and 12.5\% have ages beyond 1 Gyr.
The field star population within 6 arcmin (80 pc) radius 
is studied based on a CMD of 6919 stars. The CMD is shown in figure~\ref{figure2}.

The estimation of SFRs (figure~\ref{figure2d}) shows that the region has experienced a substantial increament in the
SFR at ages around 500 Myr, after that the rate has decreased to near zero values at about 2 Gyr.
Between 2 -- 12 Gyr, there was very little star formation in this region. In the case of younger
ages, the star formation has continued till about 10 Myr. There has been a slight increament 
in the last 50 Myr. Therefore this region has been had continued star formation 
in the last 2 Gyr period. 
The observed stellar density is shown in figure~\ref{figure2a}, along with the synthetic
one. 
The feature which is not reproduced in the synthetic CMD is the peak
in the red clump, similar to the case of region 1. The reduced $\chi^2$ value for the fit
is 0.06. 
A small enhancement in the star cluster formation which occurred between 300 Myr \,--\, 1 Gyr,
is well correlated with the high star formation rate at about 500 Myr. This region does not
show the 100 -- 300 Myr enhancement in the cluster formation, but shows a low rate
formation of clusters, which also correlates with the low SFR seen during this period.
There is one cluster which is less than 30 Myr old, and we do find an increament in the SFR
between 10 -- 50 Myr.
Thus the star formation and the cluster formation are more or less correlated in this region.

\begin{figure*}
\centering
\includegraphics[width=17cm]{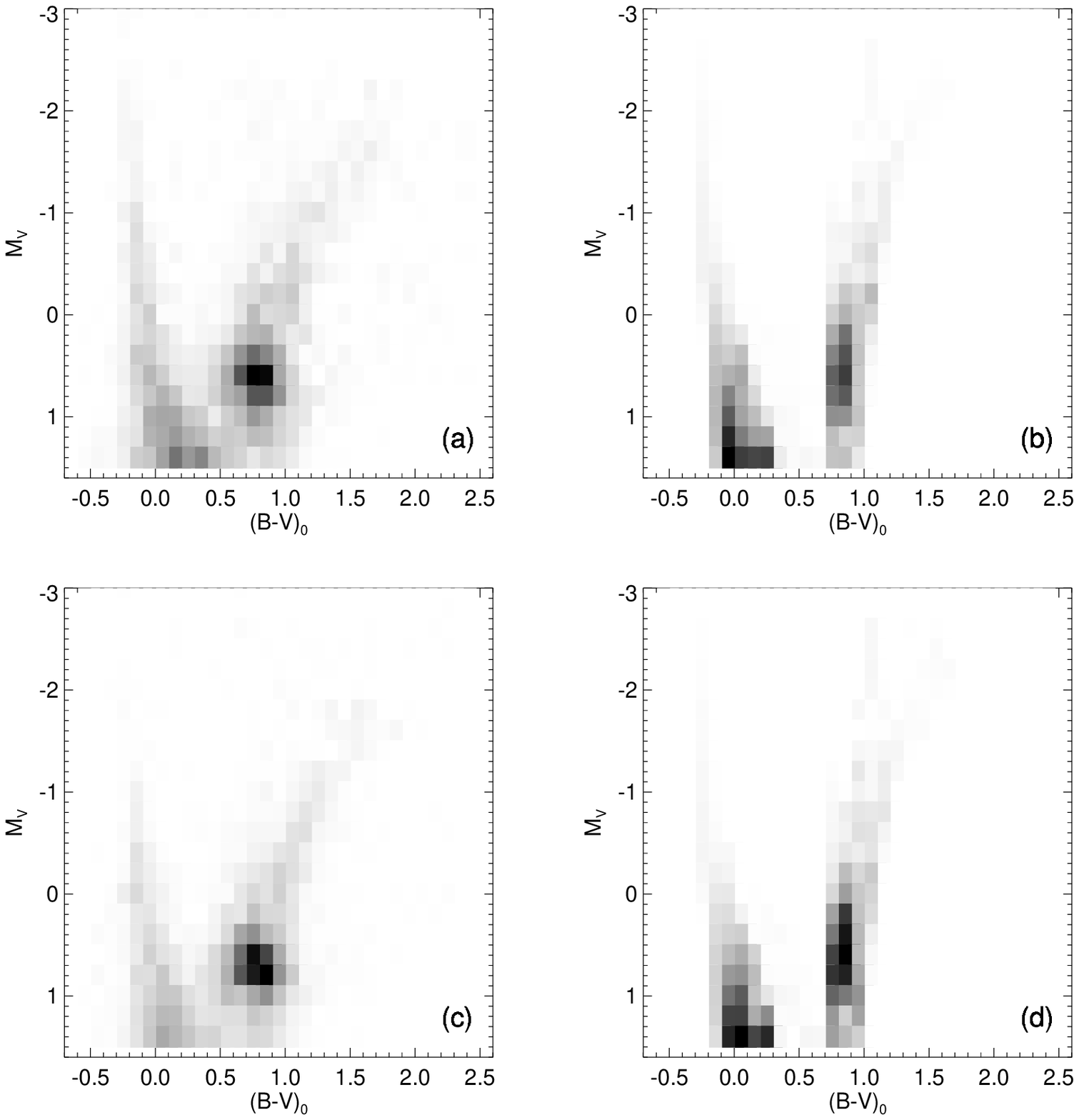}
\caption{The stellar densities on the CMDs for the regions 3 and 4 are shown here. 
The figure (a) and (c) correspond to the observed CMDs, (b) and (d) correspond to the synthetic CMDs
for the regions 3 and 4 respectively}
\label{figure4}
\end{figure*}
\subsection{Region 3}
This region is situated right at the center of the LMC Bar.
17 are found in the 30 arcmin radius, of which ages are
known for 13 clusters. Cluster formation in this region appears to be at a more
or less constant rate until 30 Myr.
31\% of clusters are with ages more than 1 Gyr, and 23\% of clusters
in the lower age ranges. Hence the cluster formation rate was lowered around 
1 Gyr, after an initial higher rate. 
The CMD of 9607 stars within a radius of 3 arcmin is shown in 
Figure \ref{figure2}. 

The SFR estimates show that this region
has increased SF between 600 -- 1200 Myr. The SFR then decreased and settled at a very
low value around 2 Gyr. During the 10 -- 12 Gyr period, this region seems to have experienced 
an enhancement in star formation. This is similar to region 1, but the estimated SFR is slightly
less. The density in the CMDs show that the simulated CMD matches well with the observed one.
The red clump peak is again not well reproduced. The reduced $\chi^2$ value of the fit is 0.05.
The observed stellar density is shown in figure~\ref{figure4}, along with the synthetic
one. 

In this region, we see that the cluster formation was more or less constant and a steady rate
of cluster formation is observed between 30 -- 1000 Myr and a slightly increased rate after
1000 Myr.
The estimated SFRs show that the star formation was not continuous between the above period, 
rather the SFR showed increament at 600 Myr, then increased further till 1200 Myr.
Then it is seen to have decreased after 2 Gyr. If we consider that the increased star
formation seen between 1 - 1.2 Gyr correlates with the increase in the cluster formation, 
then both the rates can be considered to be correlated for ages older than 1 Gyr.  Between the ages
300 -- 1000 Myr, the SFR shows increament and the cluster formation shows a slight
decreament. 
On the other hand, for ages in the range 100 -- 300 Myr, we do not find any correlation.
The cluster formation is found to have continued at the same rate, whereas the SFR is 
found to be reduced. The SFR shows an enhancement in the 50 -- 100 Myr age range, with a
decreased SFR upto the last 10 Myr. Thus the star and the cluster formation rates are found 
to be more or less correlated, except for the 100 -- 300 Myr age range.

\subsection{Region 4}
There are 24 star clusters 
in the 30 arcmin radius and ages are known for 19 of them. 
 The maximum number of star clusters were 
formed during the period 100\,--\,300 Myr, which is 58\% of the clusters. 
21\% of clusters were formed during 300\,--\,1 Gyr period, 10\% formed during 
the 30\,--\,100Myr period and the rest during the period before 1 Gyr. 
The CMD of stars within 2 arcmin radius plotted in figure \ref{figure2}
comprises of 5244 stars.

The synthetic CMD as shown in figure \ref{figure4} indicates that it
has more MS stars and a wider RGB. The reduced $\chi^2$ value for the fit is found to be 0.1,
which is slightly higher when compared to the other regions. The SFR shows peak values
between 500 Myr and  1.3 Gyr. A slight enhancement is seen around 2 Gyr. The SFR is found to be
very low in the age range 2 -- 10 Gyr, with a relatively higher rate between 10 -- 12 Gyr.
On comparing the star and cluster formation episodes, we find that
the star formation event which occurred in the age range 1.0 -- 2.0 Gyr has
managed to form two star clusters. The star formation which continued to younger
ages, upto 300 Myr, has resulted in the formation of 4 star clusters. 
The cluster formation episode is seen to be quite strong in the age range 
100 Myr \,--\, 300 yr, whereas very minimal SFR is estimated in the same age range.
Two star clusters are found to be younger than 100 Myr, an enhancement in the SFR is 
found for ages younger than 50 Myr. Thus the SFR and the cluster formation are well correlated
except for the age range 100 -- 300 Myr.

\begin{figure*}
\centering
\includegraphics[width=17cm]{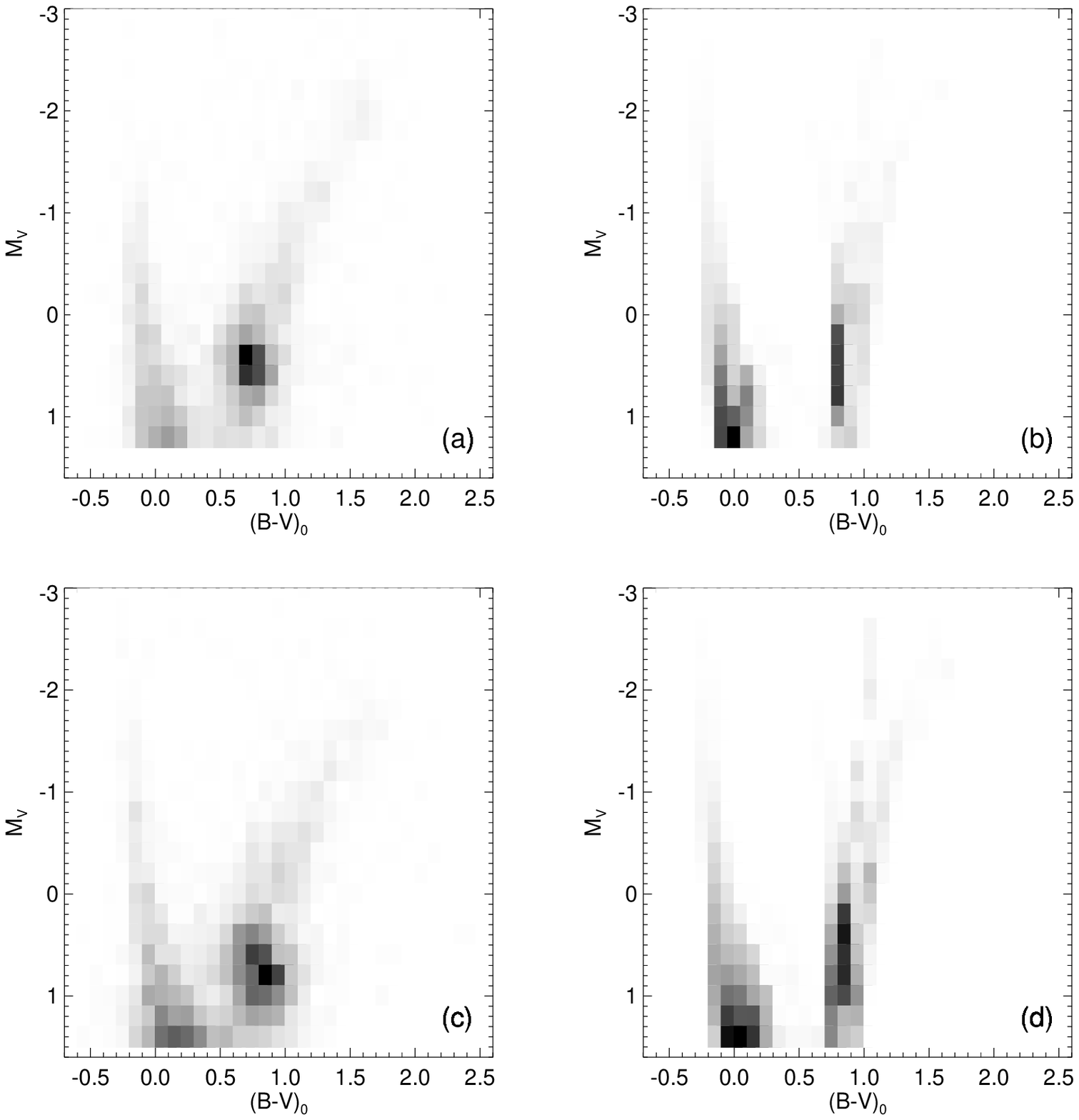}
\caption{The stellar densities on the CMDs for the regions 5 and 6 are shown here. 
The figure (a) and (c) correspond to the observed CMDs, (b) and (d) correspond to the synthetic CMDs
for the regions 5 and 6 respectively}
\label{figure6}
\end{figure*}

\subsection{Region 5}
Within a radius of 400 pc, we identified 17 star clusters and ages 
are known for 15 of them. We find that,
40 \% of the clusters have ages in the range 100\,--\,300 Myr, 33.3\% of clusters
have ages in the range 300\,--\,1 Gyr and 13.3\% clusters are older then 1 Gyr. 
The CMD of 6503 field stars within a region of 2 arcmin (27 pc) radius is
plotted in figure~\ref{figure2}.

The synthetic and the observed CMDs are compared in figure \ref{figure6}.
The synthetic CMD shows a vertical RGC and more stars at the fainter end of the MS,
which are not seen in the observed CMD. The reduced $\chi^2$ value for the fit is found 
to be 0.14 and this is the worst fit among all the regions.
The estimated SFRs show that the values are high in the age range 300 -- 800 Myr.
Very small SFR is found between 2 -- 8 Gyr and
a slightly high rate between 8 and 12 Gyr.
On comparing the cluster formation rate, the two clusters older than 1 Gyr could be formed
between 1 -- 2 Gyr along with the star formation. The enhanced cluster formation between
300 -- 1 Gyr is correlated well with the enhancement in the SFR. Six star clusters are
formed between 100 -- 300 Myr, whereas an enhancement in the SFR is found at $\sim$ 250 Myr.
One cluster is found younger than 30 Myr and one between 30 -- 100 Myr. Correspondingly,
enhancement in the SFR is found for ages younger than 50 Myr. Hence we find a more or less
correlated star and cluster formation for all ages.

\subsection{Region 6}
27 star clusters 
are found within the 30 arcmin radius, of which the ages are known for 22. 
18\% of the clusters have ages within 30\,--\,100 Myr, 50\%  have ages between 
100\,--\,300 Myr and 32\% have ages between 300\,--\,1 Gyr. These indicate that
there was a burst of cluster formation some time during 100\,--\,300 Myr with a 
tapering star formation before and after this burst. 
3302 field stars located within 4 arcmin (54 pc) radius were identified and the
CMD of the field stars are plotted in figure~\ref{figure2}. 

The synthetic CMD as shown in figure \ref{figure6} shows that the RGC population is
not matched, though the overall match could be considered satisfactory. The reduced
$\chi^2$ value is found to be 0.06.  The estimated SFRs
are found to be high in the interval of 50 Myr -- 1000 Myr, and then decreasing upto
2 Gyr. Very low SFR is found in the interval of 2 -- 10 Gyr, with a relatively higher rate
in the 10 -- 12 Gyr interval. On comparing the SF and the CF episodes,
both the rates are found to increase at about 1 Gyr. The CF rate increases further at  300 Myr,
whereas the SF rate is reduced. This region is found to show recent star formation corresponding 
to the presence of young star clusters. 4 star clusters are found in the age range 30 -- 100 Myr, 
whereas the SFR is found to decrease slightly at 50 Myr. Therefore, the SF and CF seem are found 
to be less correlated in the age range 30 -- 100 Myr, whereas indications of anti-correlation 
are found in the 100 -- 300 Myr age range.

\section{Discussion}
The aim of the present analysis is to verify the correlation between the star and 
cluster formation rates at different epochs in the LMC, particularly at smaller scales.
The present study shows that
SFRs and the cluster formation rates are more or less correlated for the age range 30 -- 100 Myr. 
In the age range 100 -- 300 Myr,
5 regions show an enhancement in the cluster formation, whereas
such an enhancement is not noticed in the SFR.  
In the age range of 300 -- 1000 Myr, the SF as well as the cluster formation shows enhancement in
their rates. 
It is also found that the SF and CF are more or less correlated for ages more than 1 Gyr.

Girardi et al. (\cite{g95}) estimated the age distribution of star clusters in LMC.
They found three periods of enhancement in the formation of star clusters in LMC,
namely at 0.1 Gyr, 1--2 Gyr and 12 -- 15 Gyr. 
Pietrzynski \& Udalski (\cite{pu00}) found peaks in star cluster formation
at 7 Myr, 125 Myr and 800 Myr. They also found peaks at 100 Myr and 160 Myr,
which they attribute to the last encounter of the Magellanic Clouds. 
Grebel et al. (\cite{g99}) found that the age distributions of both LMC and SMC
clusters peak at 100 Myr, when the Clouds had their closest encounter and last
perigalacticon.
All the above indicate that the substantial number of clusters seen in the
100 -- 300 Myr around 5 regions could be part of the cluster formation peak
at that time. 
Hence the anti-correlation between the SF and CF during this period could indicate that
the type of SF which induced the CF is different in this period.
On the other hand, the peak in the CF at 800 Myr is well correlated with the SF, where
in most of the regions, the SFR was highest around the age of 800 Myr.
 Therefore, the highest correlation between CF and SF is found at $\sim$ 800 Myr, whereas
anti-correlation is found in the age range of 100 -- 300 Myr.

It is possible that some assumptions as well as the choice of data could modulate or bias
the results obtained here. Some of the possibilities are explored below.
(1) Incompleteness in the cluster data: The incompleteness affects the fainter clusters
 and hence the older clusters will be more
affected, than the younger ones with the bright stars. This can only increase the
discrepancy observed here and not reduce it.
(2) Inappropriate estimates of the SFRs: An inspection of the CMDs of the regions presented in
figure~\ref{figure2} indicates low star
formation rates for ages younger than a few hundred Myr. In most of the cases the brightest few stars 
observed belong to 100 Myr age range. There could be bias for very bright stars as 
they could have got
saturated. This can only affect the very young population, but not for ages older than 50 Myr.
(3) Statistical significance: It is very unlikely as the discrepancy is found for 5 out
of 6 regions studied here. All the regions have good number of star clusters especially for ages
younger than 300 Myr.
(4) Selection effect: The regions studied here do not have any apriory information or property.
The choice was only based on the availability of clusters in the neighbourhood. Hence the 
results is very unlikely to be affected by this.
(5) The center field may not be the true representative of the entire 400 pc region.
 We have assumed that the stars in the central few pc could represent the radius of 400 pc.
 If this assumption is not valid such that there is difference in population within the radius,
 the results above may not be valid. To check this, the CMDs were created at for the annulus 
  between 25 and 30 arcmin, which would sample the stars near the periphery. When the central and
  the peripheral CMDs were compared, no significant changes in the stellar density were noticed.
  Special attention was given to the younger population and no noticeable 
increase was noticed in the stellar population younger than 300 Myr.
Hence we assert that the result obtained here is likely to be a true feature in the LMC, at least
near the bar regions.

In general star formation results in the formation of field stars as well as groups or clusters
of stars. Therefore, SFR is an indicator of the amount of star formation or the major star
formation events. The rate of cluster formation therefore is expected to be correlated with the
SFR.
The main result of this study is that the SF and CF are not seen to be correlated in the 100 --
300 Myr history of the LMC, whereas it seems to be well correlated for ages more as well
as less than the above range.
This is a surprising result as one expects the imprint of star
formation to be present in the field stars as well. 

The results presented here could indicate that the star formation in the LMC, has had more
preference to the formation of clusters, bound or unbound in the 100 -- 300 Myr age range. 
This results in the overabundance
of star clusters in this age range. Such a discrepancy decreases for older ages,
as the stellar evolution as well the dynamical evolution dissolves the clusters beyond the
limit of detection. This happens in the case of open clusters and not for the
blue globulars. The LMC is also seen to have a healthy population of binary clusters. 
The age distribution of the multiple clusters in the LMC shows that the largest
fraction falls in the 100 -- 300 Myr age range.
This also supports the idea of preferred formation of clusters. As the LMC is known to have a slowly 
rotating disk, the clusters born together are likely to spend considerable amount of time together.
As the tidal field of the LMC is very weak, the clusters located close by have better chance to merge
together, than to disrupt and move farther away.

The LMC is known to be interacting with our Galaxy and the Small Magellanic Cloud (SMC).
There are many studies which 
look for the signatures of possible encounters (Westerlund \cite{w97} and references 
therein, Maragoudaki \cite{m01}) and also studies 
which do simulation of the dynamics of
the interaction between the three galaxies (Fujimoto \& Murai \cite{fm84}, Gardiner \&
Noguchi \cite{gn96}, Gardiner et al. \cite{g94}). These studied have found that the LMC had
an interaction with our Galaxy at about 1.5 Gyr ago, 
with SMC at 0.2 -- 0.4 Gyr and LMC had a perigalacticon at 100 Myr. 
The SFRs which peaked around 800 Myr decreased for younger ages, while the cluster
formation increased in the 100 -- 300 Myr age range. Thus the cluster formation episode
of 100 -- 300 Myr could be correlated with the interaction with SMC or our Galaxy. SMC also shows a peak
in cluster formation between 100 -- 300 Myr (Grebel et al. \cite{g99}).
Some clusters could also have been formed due to the propagating star
formation started by the triggers. 
On the contrary, the interaction between the Clouds
does not seem to be the cause for the cluster and star formation peak at $\le$ 1000 Myr.
The result presented here indicates that the star formation induced by interaction may likely to
be biased towards group formation of stars, whereas such a bias is not found in the case of
star formation without any interaction. The formation of globular clusters or dense clusters 
is observed in the case of interacting galaxies. For example,
in the case of colliding galaxies, NGC 4038/4039
very massive clusters are found to be formed.

\begin{acknowledgements}
I thank the referee G.Pietrzynski for helpful comments.
\end{acknowledgements}

\end{document}